\documentclass[traditabstract]{aa}
\usepackage{graphicx}
\usepackage{txfonts}
\usepackage{natbib}
\usepackage{enumitem}
\bibpunct{(}{)}{;}{a}{}{,}

 \newcommand{\mics}{$\mu$m}
 \newcommand{\lums}{$\nu L _{\nu} \left(60 \mu{\rm m}\right)$}
  \newcommand{\msun}{M$_{\odot}$}
 \newcommand{\lfir}{$L _{60}$}
 
 \newcommand{\lhard}{$L _{X}$}
 \newcommand{\lbol}{$L _{bol}$}
 \newcommand{\smass}{$M _{*}$}
 \newcommand{\mbh}{$M _{BH}$}
 \newcommand{\redd}{$\lambda_{E}$}
 \newcommand{\ergs}{erg s$^{-1}$}
 \newcommand{\kms}{km s$^{-1}$}

\begin{document}

\title{The Mean Star-Forming Properties of QSO Host Galaxies}

\author{
D. J. Rosario\inst{1}
\and
B. Trakhtenbrot\inst{2}
\and
D. Lutz\inst{1}
\and
H. Netzer\inst{3}
\and
J. R. Trump\inst{4}
\and
J. D. Silverman\inst{5}
\and
M. Schramm\inst{5}
\and
E. Lusso\inst{6}
\and
S. Berta\inst{1}
\and
A. Bongiorno\inst{7}
\and
M. Brusa\inst{8}
\and
N. M. F\"orster-Schreiber \inst{1}
\and
R. Genzel\inst{1}
\and
S. Lilly \inst{9}
\and
B. Magnelli\inst{10}
\and
V. Mainieri \inst{11}
\and
R. Maiolino\inst{12,13}
\and
A. Merloni\inst{1}
\and
M. Mignoli \inst{14}
\and
R. Nordon\inst{1}
\and
P. Popesso\inst{1}
\and
M. Salvato\inst{1}
\and
P. Santini\inst{7}
\and
L.J. Tacconi\inst{1}
\and
G. Zamorani \inst{14}
}

\offprints{D. Rosario \email{rosario@mpe.mpg.de}}

\institute{ Max-Planck-Institut f\"{u}r Extraterrestrische Physik (MPE), Postfach 1312, 85741 Garching, Germany.
\and Department of Particle Physics and Astrophysics, The Weizmann Institute of Science, Rehovot 76100, Israel
\and School of Physics and Astronomy, The Raymond and Beverly Sackler Faculty of Exact Sciences, Tel Aviv University, Tel Aviv 69978, Israel
\and Department of Astronomy and Astrophyics, University of California, Santa Cruz
\and Kavli Institute for the Physics and Mathematics of the Universe (WPI), Todai Institutes for Advanced Study, the University of Tokyo, Kashiwanoha, Kashiwa, 277-8583, Japan
\and Max-Planck-Institut f\"{u}r Astronomie (MPIA), K\"{o}nigstuhl 17, D-69117 Heidelberg, Germany
\and INAF - Osservatorio Astronomico di Roma, via di Frascati 33, 00040 Monte Porzio Catone, Italy
\and Department of Physics and Astronomy, University of Bologna, Viale Berti Pichat, 6/2  Bologna
\and ETH Z\"urich, Institute for Astronomy, Wolfgang-Pauli-Strasse 27, 8093 Z\"urich, Switzerland
\and Argelander-Institut f\"{u}r Astronomie, Auf dem H\"{u}gel 71, D-53121 Bonn, Germany
\and ESO, Karl-Schwarzschild-Strasse 2, 85748 Garching,  Germany
\and Cavendish Laboratory, University of Cambridge, 19 J. J. Thomson Ave., Cambridge CB3 0HE, UK
\and Kavli Institute for Cosmology, University of Cambridge, Madingley Road, Cambridge CB3 OHA, UK
\and INAF - Astronomical Observatory of Bologna, Via Ranzani 1, I - 40127 Bologna, Italy
}

\date{Received .... ; accepted ....}
\titlerunning{SFR of BLAGN Hosts in COSMOS}


 \abstract{Quasi-stellar objects (QSOs) 
 occur in galaxies where supermassive black holes (SMBHs) are growing
 substantially through rapid accretion of gas. Many popular models of the co-evolutionary growth of galaxies and black
 holes predict that QSOs are also sites of substantial recent star formation (SF), mediated by important processes,
 such as major mergers, which rapidly transform the nature of galaxies. A detailed study of the star-forming properties of QSOs
 is a critical test of such models. We present a far-infrared Herschel/PACS study of the mean star formation rate (SFR)
 of a sample of spectroscopically observed QSOs to $z\sim2$ from the COSMOS extragalactic survey. This
 is the largest sample to date of moderately luminous QSOs (AGN luminosities that lie around the knee of
 the luminosity function) studied using uniform, deep far-infrared photometry.
 We study trends of the mean SFR
 with redshift, black hole mass, nuclear bolometric luminosity and specific accretion rate (Eddington ratio).
 To minimize systematics, we have undertaken a uniform determination of SMBH properties, as well as an analysis
 of important selection effects of spectroscopic QSO samples that influence the interpretation of SFR trends.
 We find that the mean SFRs of these QSOs are consistent with those of normal massive star-forming galaxies with a fixed
 scaling between SMBH and galaxy mass at all redshifts. No strong enhancement in SFR is found even among the most
 rapidly accreting systems, at odds with several co-evolutionary models. 
 Finally, we consider the qualitative effects on mean SFR trends from different assumptions about the SF properties of QSO hosts
 and from redshift evolution of the SMBH-galaxy relationship. While limited currently by uncertainties, valuable
 constraints on AGN-galaxy co-evolution can emerge from our approach.
  }

\maketitle

\section{Introduction}


Quasi-stellar Objects (QSOs) 
constitute the luminous end of the population of broad-line AGNs (BLAGNs), i.e. those that display broad permitted and semi-forbidden
emission lines in their spectra with FWHM of few to several thousands of \kms. The luminosity of QSOs -- they heavily
outshine their host galaxies, especially at ultra-violet (UV) and optical wavelengths -- allow them to be detected at very
large cosmological distances, and the low intrinsic obscuration they exhibit towards the nuclear engine make them
the principal laboratories used by researchers for understanding AGN accretion, environments and energetics.

The widespread existence of SMBHs in local galaxies and the tight relationships they exhibit with respect 
to the masses of their host galaxies \citep[e.g.,][]{magorrian98, tremaine02, graham07, aller07} 
suggest a close relationship between the stellar growth of galaxies and the phases of maximal growth of black holes. This
has stimulated much study into the co-evolutionary relationship between galaxies and AGNs. 
Most of the cosmic growth of super-massive black holes (SMBHs) takes place at $z=1$--$2$ in
luminous AGNs with bolometric nuclear luminosities \lbol$>10^{45}$ \ergs\ \citep{page04}. 
QSOs are the primary tracer of this population, though some studies suggest that much black hole growth 
may also occur in obscured phases that are missed in traditional
QSO samples \cite[e.g.,][]{martinez-sansigre05,polletta06,donley07,reyes08} 
or through X-ray selection \cite[e.g.,][]{daddi07b,gilli07,fiore08,fiore09,alexander11}. 
In spite of this, almost all models of AGN-galaxy co-evolution ascribe a
special role for the QSO population. For example, the popular evolutionary scenario that links elliptical galaxies to
gas-rich major mergers through a massive starburst predicts a brief period of luminous AGN activity which
is eventually visible as an optically bright QSO \citep{sanders88,granato04, hopkins08}. An important corollary is that
QSOs should be associated with the sites of current or post-starbursts. The exact relationship between QSOs and starbursts
depends on the nature and timing of the poorly constrained luminous obscured AGN phase 
believed to exist before strong feedback clears out the dust and gas from the merger remnant. However, a close
correspondence between QSOs and recent starburst events is even predicted in co-evolutionary
models that do not explicitly rely on major galaxy mergers to fuel QSOs \cite[e.g.,][]{ciotti07}. 

Star-formation (SF) in QSO host galaxies has been extensively studied using high-resolution imaging in the 
optical \cite[e.g.,][]{bahcall97,dunlop03, jahnke04} and near-infared \cite[e.g.,][]{kukula01,guyon06,veilleux09}; 
emission line tracers such as the [O II] line \citep{hes93,ho05,silverman09,kalfountzou12};
mid-infrared emission lines and PAH features \citep{netzer07,lutz08,shi09}; and far-infrared and sub-mm photometry
\citep[e.g.,][]{priddey03,omont03,lutz10,serjeant09,serjeant10,bonfield11}. In general, QSO hosts are in massive, 
spheroidally-dominated galaxies, which frequently show signs of on-going star-formation \citep[e.g.,][]{jahnke04,trump13}, 
though signatures of early stage mergers or strong disturbances are not particularly frequent 
\citep{dunlop03, guyon06, bennert08,veilleux09}. Very powerful starbursts are known to exist among high redshift
QSOs, with $\sim30$\% of very optically luminous systems showing SFRs at the level of thousands of \msun/yr 
at $z\sim2$ \citep[e.g.,][]{omont03, wang08}. Additional evidence from CO and [C II] observations indicate 
large gas supplies that could fuels starbursts \citep[e.g.][]{walter04, walter09, coppin08, wang10, wang13}. 
However, it is clear that not all QSOs are in strong starbursts -- local PGQSOs
span SFRs ranging from very low to $\sim100$ \msun/yr \citep{schweitzer06,netzer07}. 
Studies of high-redshift luminous Type II AGNs, obscured counterparts of QSOs,
suggest typically modest SFRs comparable to normal SF galaxies \citep{sturm06,mainieri11}.

Understanding the link between strong bursts of SF and QSO activity is complicated by a few important biases. Firstly, bright QSOs are
essentially all in very massive galaxies. For an evaluation of whether
QSOs are indeed in galaxies with abnormally high levels of SF, a proper comparison has to be made with the SFRs
of inactive galaxies at the same redshifts and of similar stellar mass, since SFR is strongly correlated both with redshift and
stellar mass among SF galaxies
\citep{noeske07, elbaz07, daddi07a, wuyts11, whitaker12}.
Secondly, BLAGN populations are essentially defined by spectroscopic surveys, which have important selection effects
that must be taken into account when statistically evaluating black hole and host galaxy properties 
\cite[][and Sec.~4.1]{shen12}. In this study, we explore the SF properties of QSO hosts, relying on the far-infrared (FIR) 
as a relatively clean measure of the total luminosity of SF-heated dust \citep{netzer07, rosario12}. We start by
compiling one of the largest samples of broad-line AGNs in a deep extragalactic survey field with uniformly measured
SMBH properties. The combination of sample size, redshift coverage, and spectroscopic and FIR imaging depth is
unsurpassed in existing studies of distant QSOs. From this compilation, we determine the mean SFRs of 
moderately luminous QSOs through the stacking of Herschel/PACS images, while using simple models to account
for the effects of sample biases and explore the relationship between SMBH growth and global star-formation.

As the sample consists of fairly luminous systems, we use the term ``QSOs" or ``BLAGNs" to refer to 
all broad-line AGNs throughout this paper. Additionally, in Section 5.1, we compare our QSOs with 
AGNs selected using X-rays, which comprise a much larger and diverse set of objects. Such
``X-ray AGNs" encompass broad-line, narrow-line or optically dull AGNs.

We assume a standard $\Lambda$-CDM Concordance cosmology, with $H_{0} = 70$ \kms~Mpc$^{-1}$ and $\Omega_{\Lambda}=0.7$. Stellar masses in this study, where reported assume a Chabrier Initial Mass Function \citep{chabrier03}. 

\section{Datasets and Sample Selection}

\subsection{Selection of Broad-Line AGNs}

For a substantial sample of BLAGNs over a range of redshifts with associated deep far-IR and X-ray imaging
coverage, we concentrated on the 2 deg$^2$ medium-deep COSMOS extragalactic survey field \citep{scoville07}. This
field has been the target of multiple optical spectroscopic surveys of varying depths and we turn to several of these datasets to select
BLAGNs. In practice, the size of the sample is restricted to the redshifts at which broad H$\beta$ and MgII $\lambda2800$ lines are
reliably measured in optical spectra, as well as the capabilities of the broad-line fitting method used to derive SMBH masses. 
For example, several sources were excluded after a manual inspection of their fits, either because they suffered from bad data,
were too close to the edge of a spectrum or were simply limited by the S/N of the spectra.

QSOs were selected from the Sloan Digital Sky Survey (SDSS) DR7 spectroscopic database \citep{abazajian09} which
covers the COSMOS field. Targets with high quality redshifts and classified as `QSO' were identified, from which radio-loud
sources (based on 20 cm fluxes in the FIRST survey) and broad absorption line systems (BALs) were removed. Details
of the selection method can be found in \citet{trakhtenbrot12}. Our total SDSS subsample consists of 70 BLAGNs with
measurable SMBH masses.

The zCOSMOS survey \citep{lilly07, lilly09} is a multi-purpose spectroscopic campaign in COSMOS that uses the VIMOS spectrograph on the VLT.
It consists of two tiers: zCOSMOS-Bright targets 20000 galaxies over the entire COSMOS/ACS field to $I_{AB} = 22.5$, while zCOSMOS-Deep
targets 10000 galaxies over the inner 1 deg$^2$ of the field to a deeper limit of $B_{AB} = 25$ with an additional color-based
galaxy preselection. From both tiers, QSOs were classified based on spectral features and an automated comparison to a QSO template,
followed by visual assessment of the fit. Only objects with the Mg II $\lambda2800$ line were included in our sample. z-COSMOS yields a total of 176 BLAGNs with measurable SMBH masses, 146 from the Bright survey and 30 from the Deep survey.

The XMM-COSMOS X-ray survey \citep{cappelluti09} has yielded an extensive X-ray point source catalog in the COSMOS field \citep{brusa10},
which has, in turn, been the basis of spectroscopic follow-up programs. In addition to the SDSS and zCOSMOS sources described
above, we incorporated a sample of BLAGNs selected from a Magellan/IMACS program of X-ray source follow-up \citep{trump07, trump09a} which yielded 112 BLAGNs for which we could measure SMBH masses. Nominal flux limits for this dataset are
$i^{+}_{AB} = 23.5$, sampling fainter sources than zCOSMOS-Bright. 

Combining the three subsamples and resolving duplicates (i.e, the same AGN observed in two or more spectroscopic surveys), we
arrived at a final QSO ``working sample" of 289 objects. 
This is currently the largest sample of BLAGNs with deep, uniform FIR coverage and reliably estimated SMBH properties.

\subsection{Herschel Imaging and Photometry}

The FIR data used in this work were collected by the PACS instrument \citep{poglitsch10} on board the Herschel Space 
Observatory, as part of the PACS Evolutionary Probe \citep[PEP,][]{lutz11} survey. Observations of almost the entire 
COSMOS field were taken in two PACS bands (100 and 160 \mics).
We make use of PACS catalogs extracted using 
the prior positions and fluxes of sources detected in deep MIPS 24 \mics\ imaging in the field \citep{lefloch09}. 
This allows us to accurately deblend PACS sources in images characterized by a large PSF, 
especially in crowded fields, and greatly improve the completeness of faint sources at the detection limit. 
The 3$\sigma$ limits of the PACS catalogs are 5.0/11.0 mJy at 100/160 \mics. 
We consider sources below these limits as undetected by PACS. 
Residual maps, from which all detected sources are subtracted, were used in the stacking procedure described in Section 3.2.
Detailed information on the PEP survey, observed fields, data processing and source extraction may be found in \cite{lutz11}.

Of the 272 QSOs that lie within the region of uniform FIR coverage, 
38 (14\%) were detected in both PACS band. This is slightly higher than the FIR detection rate of 10\% among 
more luminous QSOs presented in \citet{dai12} from somewhat shallower Herschel/SPIRE imaging. Given that
their bolometric luminosity and redshift ranges are different from ours, we refrain from a detailed
comparison of the two samples, but note that the rough consistency in the detection rates
suggests that the FIR luminosity of QSOs does not rise dramatically with nuclear luminosity.

\subsection{X-ray Photometry}

We crossmatched our working sample to the optical counterparts of X-ray point sources from the XMM-COSMOS survey 
\citep{cappelluti09, brusa10}. 243 BLAGNs have an associated X-ray point source, which is 86\% of the sample. Of the remaining
39 sources with no X-ray counterpart, a fraction lie on the edges of the XMM-COSMOS field, where the X-ray depths
are shallower than in the center of the field. However, there are a few genuine BLAGNs that have no X-ray counterparts 
even at the center of the XMM-COSMOS field. These are examples of relatively rare X-ray faint but optically luminous AGNs
\citep{vignali01}.

We make use of absorption-corrected rest-frame X-ray luminosities in the hard band (2-10 keV; \lhard\ hereafter) for X-ray
AGNs and QSOs with X-ray detections. Details of the estimation of \lhard, as well as information about redshifts and
other properties of the XMM-COSMOS catalog used here, may be found in \citet{rosario12} and \citet{santini12}.

\section{Methods}

\subsection{SMBH masses and bolometric luminosities}

\begin{figure*}[t]
\includegraphics[width=\textwidth]{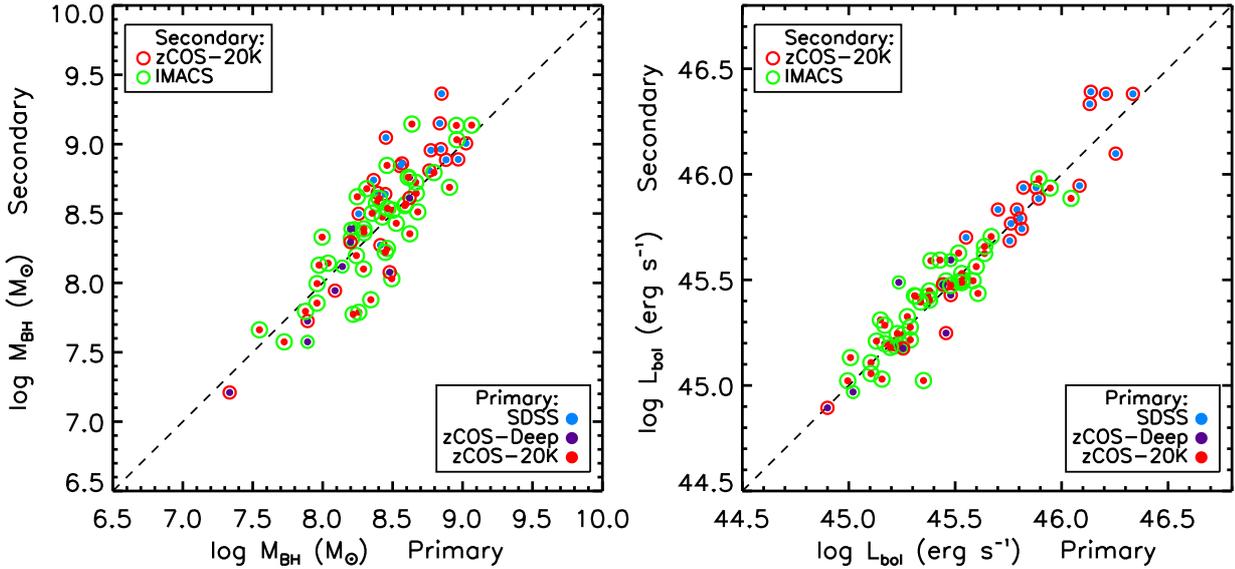}
\caption{ A comparison of black hole mass (\mbh) estimates (left panel) and bolometric luminosity (\lbol) estimates (right panel)
for a set of QSOs that have at least two spectra from different datasets. The estimate from the
``primary'' dataset is plotted on the X-axis and the estimate from the ``secondary" dataset is plotted on the Y-axis, where the
hierarchy is based on preferences outlined in Section 4. The
color of the solid core point identifies the primary dataset, which is either SDSS (blue), zCOSMOS-Deep (purple) or 
zCOSMOS-Bright (red). The color of the open encircling point identifies the secondary dataset, which may be zCOSMOS-Bright (red) or
IMACS (green). The \lbol\ estimates, derived from direct measurements of the local continuum from the spectra, 
are very consistent between datasets. The \mbh\ estimates show more scatter and a small systematic variation about
the 1:1 line.
}
\label{overlaps}
\end{figure*}

Before embarking on the measurement of SMBH masses, we checked the relative spectrophotometric
performance of the zCOSMOS and IMACS datasets by comparing the spectra of a set of objects that overlapped 
between these two samples. We found systematic flux offsets between the spectra at the level of $\approx0.4$ dex.
The comparison of zCOSMOS and SDSS spectra of a small set of overlapping objects also suggested an offset
of $\approx0.3$ dex. The IMACS and zCOSMOS spectra are subject to a seeing-dependent slit-loss due to the 1" width of the
slit \citep{trump09a, merloni10}, which could be the cause for most of the observed offsets.

To account for these remaining spectrophotometric offsets, we measured synthetic broad-band magnitudes
directly from the zCOSMOS and IMACS spectra and compared them to integrated broad-band photometry of the QSOs 
in the public COSMOS multiwavelength catalog \citep{capak07}. The differences were used to estimate correction factors 
that were then applied to the spectra. The zCOSMOS-Bright and IMACS spectra, which cover the approximate
wavelength range of 5500--9500 \AA, were scaled to CFHT i* photometry, while the zCOSMOS-Deep spectra, with wavelength
coverage from 3600--6800 \AA, were scaled to Subaru g+ photometry. A visual comparison of the spectra of overlapping
objects after applying the scaling factors showed broad consistency in both the continuum normalization and the shape of
the spectra.

Black hole masses (\mbh) were estimated using virial relationships calibrated from the reverberation mapping
of local BLAGNs, following the methodology detailed in \cite{trakhtenbrot12}. We only used virial relationships
for the broad H$\beta$ and MgII $\lambda2800$ emission lines, which effectively restricts the redshifts probed
by our sample to $z<2.2$. It has been proposed that, in principle, the broad CIV$\lambda1550$ line can be 
used to estimate \mbh\ at higher redshifts. However, several studies have shown that such 
estimates are highly unreliable see \citep[see][and references therein]{trakhtenbrot12} and we therefore
exclude CIV-based masses in this work.

Our fitting method is detailed in \cite{trakhtenbrot12}. Broad lines are modeled as combinations of 
broad and narrow gaussian components fit along with the underlying continuum, narrow absorption 
features and with an optimised template to account for FeII and FeIII band emission.
All the fits were visually inspected and vetted, while a fraction were improved manually.

AGN bolometric luminosities (\lbol) were estimated using bolometric corrections to the monochromatic
luminosities at either 5100 \AA\ or 3000 \AA\ rest-frame. The choices of bolometric corrections
are derived in \cite{trakhtenbrot12} and are consistent with the prescriptions of \citet{marconi04}, 
though slightly lower than some other commonly-used values \citep[e.g.,][]{richards06}. 

We calibrate the performance and uncertainties on our measurements by comparing \mbh\ and \lbol\
for a set of 63 QSOs with spectra in two or more datasets. These comparisons are shown in Figure \ref{overlaps}.
The quality of the relative spectrophotometric calibration governs the rms scatter of $\approx0.11$ dex in the 
independent measurements of \lbol. The scatter in \mbh\ is a bit larger ($\approx0.24$ dex) and reflects the sensitivity
of the broadline fitting technique to the S/N of the spectra and the effects of fringing noise in the red ends of the zCOSMOS spectra.
In particular, we find a significant offset of $\approx 0.2$ dex between \mbh\ from SDSS and zCOSMOS spectra, towards higher masses
from the latter dataset -- inspection suggests that this is due to fringing in the red part of the zCOSMOS spectra affecting the fits. 
For this reason, we default to using the SDSS fits for objects where an overlap exists between the two datasets. We adopt 
a conservative uncertainty of 0.3 dex in \mbh\ in further analysis, which takes into account the scatter and possible systematic
offsets across the fits. 

\subsection{FIR luminosities: detection, stacking and measurement}

As a direct tracer of the FIR emission we concentrate on the mean luminosity \lums, estimated at a rest-frame wavelength of 
60 \mics (henceforth, \lfir). This choice is set by a wavelength long enough to avoid significant
AGN contamination and short enough to be sampled by PACS 160 \mics\ observations even at the 
highest redshifts considered in this work. 

We study trends of \lfir\ of BLAGNs binned in redshift and additionally in SMBH mass (\mbh), 
AGN bolometric luminosity (\lbol) or SMBH specific accretion rate, expressed as the fraction of the Eddington luminosity (\redd).
\lfir\ for objects in a bin were measured from our Herschel PACS data in a manner detailed in \citet{shao10} and \citet{santini12}.
We briefly describe it here.

At each PACS band, a small fraction of sources ($\approx 10{\rm \% -}15$\%) are detected in both PACS bands. \lfir\ is
calculated for these using their individual redshifts and a log-linear interpolation of PACS fluxes. Of the remaining sources,
some are detected in only one PACS band, while the majority are undetected in the FIR data. For the latter, we stacked
at the optical positions of the AGNs on PACS residual maps using routines developed on the \cite{bethermin10} FIR stacking libraries,
from which we derive mean fluxes in both bands using PSF photometry. We then average the stacked fluxes with the fluxes of
sources singly detected in either PACS band, weighting by the number of sources. This gives mean fluxes for the 
partially detected and undetected AGNs in both bands, from which we derive a mean \lfir\ using the median redshift of 
these sources. The final 60 \mics\ luminosity in each bin was computed by averaging over the linear luminosities of 
detections and non-detections, weighted by the number of sources. This procedure was only performed for bins 
with more than 3 sources in total.

Errors on the infrared luminosity are obtained by bootstrapping, in a fashion similar to that used in \cite{shao10}. A set of sources
equal to the number of sources per bin is randomly chosen 100 times among detections and non-detections (allowing repetitions), and
 \lfir\ is computed per each iteration. The standard deviation of the obtained \lfir\ values gives the error on the average 
 60 \mics\ luminosity in each bin. The error bars thus account for both measurement errors and the scatter in the population distribution. 
 
\subsubsection{AGN contamination in the FIR}

Most of the AGNs studied in this work are relatively luminous systems, spanning the turnover in the 
AGN luminosity function \citep[e.g.,][]{marconi04, hopkins07}. Even if a fraction of their
bolometric output is reprocessed by cold dust in their host galaxies, these
AGNs could significantly alter, or even dominate, the FIR luminosity of their host galaxies. However,
several studies of QSOs have shown that most of the dust reprocessed output in AGNs is in the form
of hot dust emission, which peaks at mid-IR wavelengths \citep{schweitzer06, netzer07, rafferty11, mullaney11, mor12, rosario12}
and drops off steeply to the FIR. This implies that the 
AGN bolometric correction at a rest-wavelength of 60 \mics\ $k_{60}$ ($\equiv L_{bol}/L_{60}$) is $\gg 1$.  
Accounting for the diversity of empirically determined AGN SEDs \citep{netzer07, mullaney11}, 
and using the methodology described in \cite{rosario12}, we estimate the bolometric correction 
to have the following form:

\begin{equation}
\log k_{60} \; = \; 1.65 + 0.2 \log L_{bol,46}
\end{equation}

\noindent where $L_{bol,46}$ is \lbol\ in units of $10^{46}$ \ergs. The expected scatter in $k_{60}$
is $\approx 0.3$ dex, due to intrinsic variation in the IR SED shapes of AGNs and from the real
scatter in the local X-ray to MIR correlation used in \citet{rosario12} to connect IR to total AGN emission. 

\begin{figure*}[t]
\includegraphics[width=\textwidth]{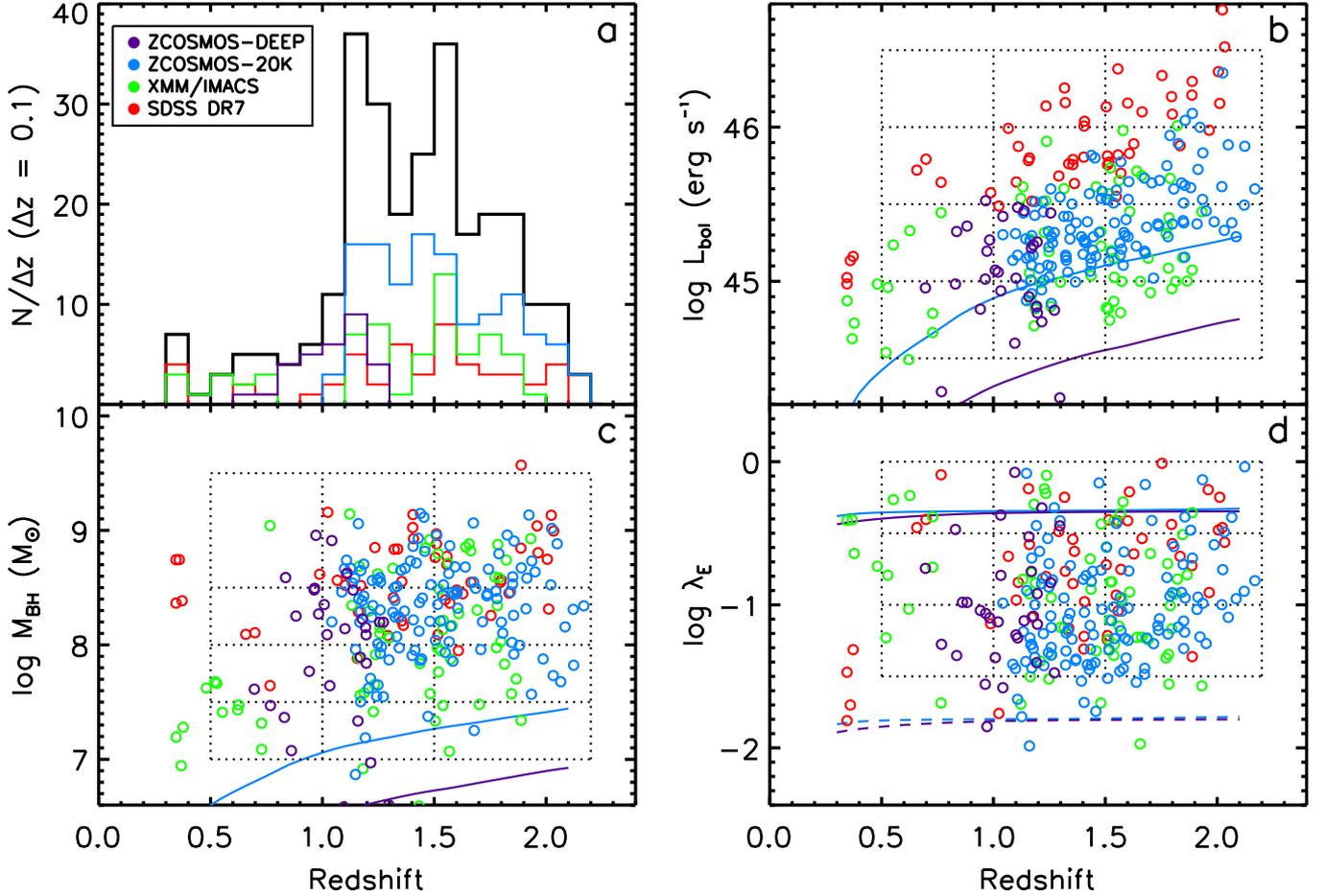}
\caption{ Properties of the COSMOS QSO working sample (duplicates resolved). 
(a) The redshift distribution of the sample: all QSOs (black) and for each individual
dataset, colored according to the key in the upper left corner. 
(b) AGN bolometric luminosity against redshift. The solid lines show the limiting \lbol\ set by the $I_{AB}=22.5$ and
$B_{AB}=25$ depths of the zCOSMOS Bright and Deep surveys respectively. 
(c) Black hole mass (\mbh) against redshift. The solid lines show the limiting \mbh\ set by the limiting magnitude of the 
zCOSMOS datasets for SMBHs with broad line FWHM$=1500$ \kms. (d) Eddington ratio (\redd) against redshift.
Lines show the \redd\ limits set by the limiting magnitude of the zCOSMOS datasets, for two different
broad line FWHM: 1500 \kms\ (solid) and 8000 \kms\ (dashed). The points and limit lines in the all panels
are colored by dataset according to the key in panel (a). The bins used to divide the sample by redshift and \mbh, \lbol\ and \redd\
are shown as dotted lines in the respective panels. Objects in each binned subsample 
were stacked together in the Herschel/PACS maps to derive mean FIR luminosities for the bin.
}
\label{sample_props}
\end{figure*}

\section{Sample Properties}

\begin{figure*}[t]
\includegraphics[width=\textwidth]{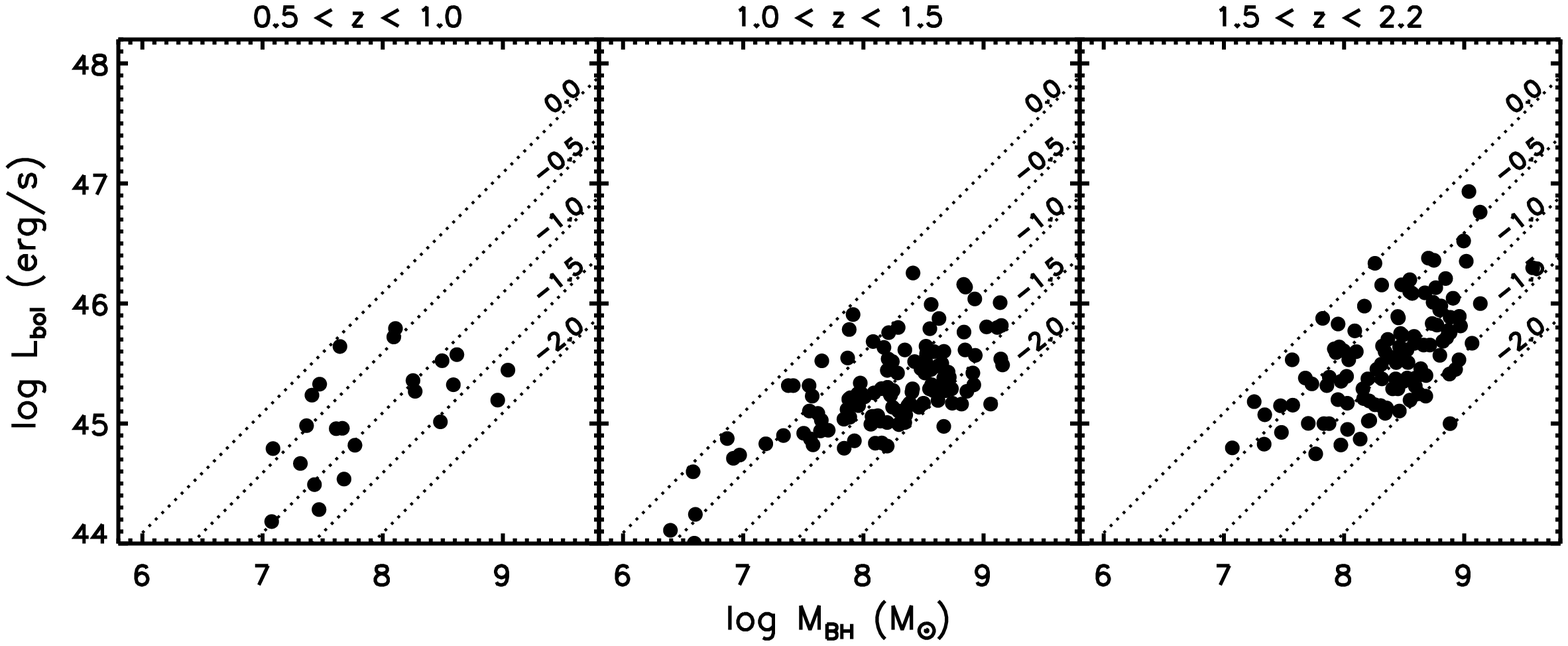}
\caption{ Black Hole mass (\mbh) vs. Bolometric luminosity (\lbol) for QSOs in three redshift bins
in the COSMOS field. Lines of constant Eddington ratio (\redd) are shown as dotted sloped lines and labelled by log \redd.
}
\label{mbh_lbol}
\end{figure*}

The redshift distribution of our BLAGN sample is shown in Figure \ref{sample_props}a. Most of the sample lies between
$z=1$ and $z=2$, because the zCOSMOS AGN, the largest part of the sample, were specifically chosen to include the MgII line, 
which enters into the wavelength range of the zCOSMOS-Bright spectra at $z\sim1$. The SDSS, IMACS and zCOSMOS-Deep 
subsamples (the latter covering bluer wavelengths than zCOSMOS-Bright) contribute to the tail of sources at lower redshifts.

For all the FIR stacking analyses in this work, we divide the AGNs into subsamples on the basis of redshift. We will use the
following fiducial redshift bins: $0.5<z<1.0$, $1.0<z<1.5$, $1.5<z<2.2$ -- designated as the `low', `intermediate'
and `high' redshift bins respectively.

In the other three panels of Figure \ref{sample_props}, we plot the the SMBH mass (\mbh), the AGN bolometric luminosity (\lbol)
and the specific accretion rate or Eddington ratio (\redd), against redshift.
The bins in redshift, \mbh, \lbol\ and \redd\ used in the stacking analyses of Section 5 are 
shown as boxes with dotted outlines in the Figure. These bins cover essentially all the AGN in our sample. 
Details of the binning scheme, including the number of objects in each bin, are listed in Table 1.

In the Figure, different colors are used to represent the different spectral datasets used in this work. 
In cases where the same object is observed in two different programs, we
adopt one of the measurements following the hierarchy SDSS $>$ zCOSMOS-Deep $>$ zCOSMOS-Bright $>$ IMACS
(justified in Section 3.1). 

AGN from the different spectral surveys occupy different and complementary areas in the space of BLAGN properties.
The sharpest contrast is in \lbol, where the SDSS AGN occupy the luminous end at all redshifts, while the
zCOSMOS and IMACS AGN fill in the sample to lower luminosities, providing a roughly uniform sampling of AGN with 
\lbol$> 10^{44.5-45.0}$ \ergs\ over all redshifts. SDSS AGN are also typically at higher \mbh\ and higher \redd\ than AGN
from the other surveys. The zCOSMOS-Deep AGN occupy a fairly narrow range in redshift between $0.8<z<1.3$ and
a range of \lbol\ which is lower than AGN from the other samples. The parent zCOSMOS-Deep spectral survey has
a color-preselection which chooses galaxies with $z>1.5$, which is in contrast to the actual redshifts of the AGN
from the survey. This is likely because the zCOSMOS-Deep AGN have optical SEDs that are dominated by AGN
light, while the color-preselection is only valid for galaxy-dominated SEDs. 

The flux limits set by the noise properties of the spectral datasets or the magnitude limits of the various surveys 
introduce redshift-dependent luminosity limits to our sample. For e.g., the limits of the two zCOSMOS surveys
are shown as solid lines in Panel b. The IMACS spectra go to fainter fluxes than zCOSMOS-Bright at $z>1.5$, 
but still describe a flux-limited subsample. The limits shown here have been calculated for MgII $\lambda2800$, 
the primary broad line used for mass measurements in this work. At $z\lesssim0.7$, only H$\beta$ is visible
in most spectra, as MgII enters the UV atmospheric cutoff. However, since our method is cross-calibrated
between these two lines, the limits shown here are essentially identical irrespective of the line used for the \mbh\ measurement.

There is an equivalent lower limit to \mbh, since low mass SMBHs produce ``narrow" broad-lines, with FWHM$<1500$ \kms,
which will not be easily identified as BLAGNs in spectral datasets. This, combined with the dependence of \mbh\ on \lbol,
sets a approximate lower envelope to the SMBH masses in our sample, shown with the solid lines in Panel c. There is no
formal upper-limit to the \mbh\ distribution, but SMBHs with very broad lines (FWHM$>10^5$ \kms) are never found \citep{trump09b}.
The combination of the lower limits to \lbol\ and \mbh\ sets a lower limit to \redd\ which is FWHM dependent, as 
shown for the zCOSMOS spectra in Panel d. Both ZCOSMOS Deep and Bright limits overlap in this plot.
Interestingly, the limit is almost flat with redshift. In other words, our sample selects objects with the same range in 
specific accretion rates across all redshifts in this study.

It is worthwhile noting here that the AGNs in the low redshift bin do not include many objects with high \mbh\ when compared
to the two higher redshift subsamples. This may be purely stochastic or due to Malmquist bias, 
but it does lead to some complexity in the interpretation of SFR trends in Section 5.

\subsection{Selection Effects in the \mbh\ -- \lbol\ plane}

In Figure \ref{mbh_lbol}, we plot the AGN bolometric luminosity against the SMBH mass for the QSO sample, with
separate panels for each fiducial redshift bin. Lines of constant Eddington ratio are shown as dotted lines and labelled accordingly.
The distribution of objects in this diagram has a characteristic form. In each bin, there is
a rough lower limit to \lbol, below which essentially no objects are found. This is an observational bias set by the flux
limits in our sample (see above). Besides this, one may also notice that the range in \mbh\ displayed by BLAGNs
is also a function of \lbol. In particular, at low bolometric luminosities, one may notice a tail of low mass AGNs, which are
typically absent at higher AGN luminosities. This selection effect is driven mostly by the steep drop in the incidence of AGN 
with increasing SMBH specific accretion rate \citep[e.g.,][]{schulze10}. AGNs with low SMBH masses are selected only if they have
\redd\ high enough to lie above the luminosity threshold set by the observational limits. However, 
the space density of such rapidly accreting systems is low at all redshifts, and hence most of the low mass SMBHs in our sample 
are found close to the \lbol\ limit in all three bins. 
At higher SMBH masses, even systems with low specific accretion rates satisfy the luminosity threshold. As relatively more AGNs
are found at low \redd\ than high \redd, the observed specific
accretion rate distribution of our sample will change across \mbh. When considering AGN with low mass SMBHs, our sample
contains a large fraction of rapidly accreting systems (\redd$\sim1$), while at high \mbh, one finds a majority of
slowly accreting AGN (\redd$\sim 0.03$). These patterns mirror those found among the brighter SDSS QSO population
\citep{steinhardt10}. Since \mbh\ correlates with stellar mass, which in turn, correlates with star-formation rate,
such selection effects must be kept in mind when interpreting trends between SFR and various BLAGN properties, as we
do in Section 5.

\section{Results: The Mean SFR of BLAGNs}

\begin{table*}
\caption{Mean Rest-frame 60 $\mu$m luminosities in bins of Redshift and SMBH properties.}              
\label{table1}      
\centering                                      
\begin{tabular}{c c c c c}          
\hline\hline                        
Redshift bins & 0.3 -- 0.5 & 0.5 -- 1.0 & 1.0 -- 1.5 & 1.5 -- 2.2 \\    
\hline                                   
\hline  
 All AGNs & 44.12 -- 44.36  (3/7) & 44.48 -- 44.82  (2/20) & 45.02 -- 45.14  (18/115) & 45.29 -- 45.43  (13/110) \\
\hline                                  
\hline                                 
Bolometric Luminosity (log \lbol) &  & & &   \\
\hline                                   
44.5 -- 45.0 & --- & 43.92 -- 44.27  (0/7) & 43.62 -- 44.44  (0/16) & 44.48 -- 44.90  (0/10) \\
45.0 -- 45.5 & --- & 44.65 -- 45.02  (1/8) & 44.97 -- 45.17  (10/66) & 45.06 -- 45.31  (4/43) \\
45.5 -- 46.0 & --- & 44.55 -- 45.00  (1/5) & 45.19 -- 45.38  (7/28) & 45.39 -- 45.61  (8/43) \\
46.0 -- 47.0 & --- &          ---  (0)             & 44.25 -- 44.94  (1/5)   & 45.42 -- 45.62  (1/14) \\ 
\hline                                 
\hline                                 
Black Hole Mass (log \mbh) &  & & &   \\
\hline                                   
7.0 -- 7.5 & --- & 44.23 -- 45.12  (1/5) & 45.06 -- 45.82  (1/4)  &  44.93 -- 45.29  (0/6)  \\
7.5 -- 8.0 & --- & 43.99 -- 44.47  (0/5) & 44.61 -- 44.97  (3/26) & 44.86 -- 45.32  (1/19) \\
8.0 -- 8.5 & --- & 44.30 -- 44.81  (1/6) & 44.90 -- 45.16  (3/41) &  45.01 -- 45.22  (4/42) \\
8.5 -- 9.5 & --- & 44.74 -- 44.93  (0/4) & 45.05 -- 45.23  (11/44) &  45.50 -- 45.71  (8/43)  \\ 
\hline                                          
\hline                                 
Eddington Ratio (log \redd) &  & & &   \\
\hline                                          
-1.5 -- -1.0 & --- & 44.34 -- 44.69  (0/6) &  44.90 -- 45.08  (6/53) & 45.32 -- 45.55  (7/44) \\
-1.0 -- -0.5 & --- & 44.37 -- 44.44  (0/4) &  45.07 -- 45.25  (8/39) & 45.23 -- 45.41  (4/40) \\
-0.5 -- 0.0  & --- & 44.61 -- 45.08  (2/7) &  44.51 -- 45.44  (1/11) & 45.17 -- 45.44  (2/21) \\
\hline
\hline
\end{tabular}
\tablefoot{ 
Mean Rest-frame 60 $\mu$m luminosities are in units of log \ergs.  
In parentheses at each entry: number of QSOs detected in PACS in a bin/total number of QSOs in a bin.
}
\end{table*}

\subsection{Trends with Redshift: A comparison to X-ray AGNs}

The BLAGN sample in this work is a subset of the population of luminous AGNs in the COSMOS field, specifically those
with unobscured lines of sight to the broad line region around the accreting black holes. By and large, they are also
a subset of the X-ray AGN population in that field, since only a small fraction of the BLAGNs are not detected in the XMM-COSMOS
survey (Section 2.3). In an earlier study \citep{rosario12}, we
constrained the mean SF properties, as measured by \lfir, of a larger and more complete set of AGNs
from XMM-COSMOS selected on the basis of their X-ray emission \citep{cappelluti09, brusa10}. Here, we consider
the redshift evolution of \lfir\ of our BLAGNs. This serves two purposes: it allows us to set a baseline for the SF properties
of the QSO population selected in COSMOS across redshift, as well as  to compare the typical SF properties of BLAGNs
with the larger X-ray selected population, which can be instructive in revealing potential differences between the host galaxies of QSOs
and other AGNs, as well as highlight selection biases in BLAGN samples.

\begin{figure}[t]
\includegraphics[width=\columnwidth]{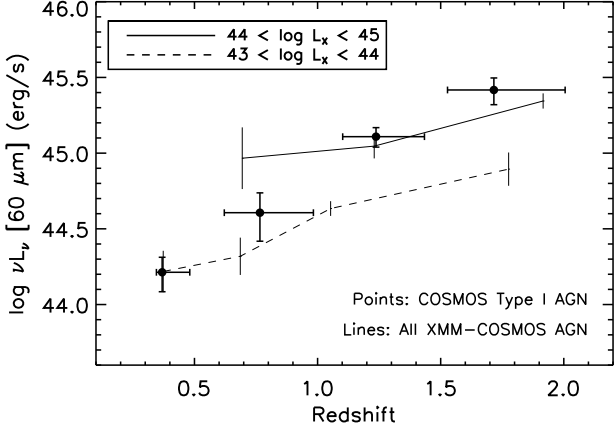}
\caption{The mean 60 \mics\ luminosity \lfir\ of QSOs in COSMOS as a function of redshift, shown as solid circular points with error bars.
The X-axis error bars show the range in redshift that encompass 80\%\ of the sample in the respective
redshift bin. The solid and dashed lines show the mean trends for AGNs from the XMM-COSMOS survey \citep{brusa10}
in two bins in instrinsic 2-10 keV X-ray luminosity. Vertical error bars placed at intervals on these lines show the $1\sigma$ scatter
in these trends.
}
\label{z_trends}
\end{figure}

In Figure \ref{z_trends}, we plot the mean \lfir\ of all BLAGNs in our sample binned by redshift. In addition to the fiducial redshift
bins listed in Section 4, we include an additional low redshift bin at $0.3<z<0.5$ for a longer redshift baseline. The x-axis error bars
show the range in redshift that contain 80\% of all objects in a bin contributing to the mean \lfir\ measurement, while
the y-axis errors come from bootstrap resampling into the stacked sample. In the Figure, we also show the evolution with redshift of the
mean \lfir\ for X-ray AGNs from \citet{rosario12} in two bins in hard-band (2-10 keV) X-ray luminosity: 
$10^{43}$ --  $10^{44}$ \ergs\ and  $10^{44}$ --  $10^{45}$ \ergs. These correspond roughly to the 
luminosities of local powerful Seyferts and QSOs respectively. A detailed discussion of the offsets in \lfir\ between these 
lines and their evolution with redshift may be found in \citet{rosario12}.

At low redshifts, the \lfir\ of the BLAGNs are consistent with that of X-ray AGNs in the luminous Seyfert range, 
while at $z>1$, their \lfir\ are comparable with the more luminous X-ray AGNs. At first glance, one may mistakenly attribute
the difference in the typical FIR luminosities of low and high redshift BLAGNs as a sign that the mean 
SFR of BLAGNs evolves more rapidly with redshift than the mean SFR of the general population of X-ray 
AGNs. However, in reality, the difference in redshift evolution is primarily governed by the change in the typical luminosity of 
BLAGNs with redshift.

\begin{figure}[t]
\includegraphics[width=\columnwidth]{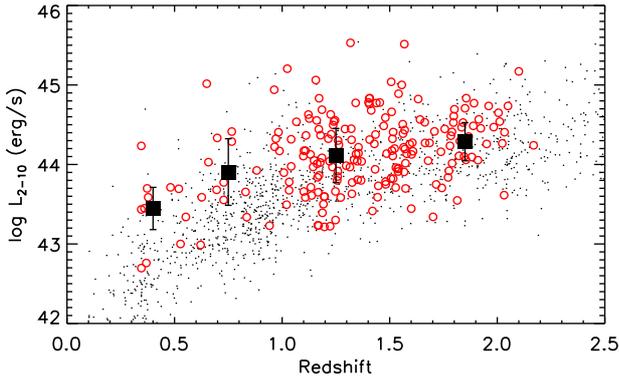}
\caption{2-10 keV luminosities of X-ray AGNs (small black points) and X-ray detected QSOs (large red open points)
over redshift, based on photometry from the XMM-COSMOS survey \citep{brusa10}. The mean X-ray luminosity
(in log units) of QSOs in our nominal redshift bins are plotted as large black square points, with the error bars showing the
median absolute deviation of the data points from the mean. QSOs are typically more luminous
than most X-ray AGNs at low redshifts, but span a similar range in X-ray luminosities as XMM-COSMOS AGNs at $z>1$.
Despite this, there is an increase in the mean X-ray luminosity with redshift for the QSOs in our sample.}
\label{lx_vs_z}
\end{figure}

In Figure \ref{lx_vs_z}, we compare the X-ray luminosities of X-ray detected BLAGNs with the 
full population of X-ray AGNs from
XMM-COSMOS. At $z<1$, BLAGNs are more luminous than the typical X-ray AGN, but at higher redshifts, both sets span
similar ranges in X-ray luminosity. Nevertheless, the typical X-ray luminosity of BLAGNs increases with redshift, from around
$10^{43.5}$ in the $0.3<z<0.5$ bin to $10^{44.3}$ at $1.5<z<2.2$. This behavior is set purely 
by the particular flux limits of the COSMOS spectral datasets.
Putting together Figure \ref{z_trends} and Figure \ref{lx_vs_z}, we see that the steeper increase of \lfir\ with redshift 
of the BLAGNs is simply due to the change in the typical AGN luminosity of the population. The mean SFR
of BLAGNs is consistent with the mean SFR of similarly luminous X-ray AGNs. 

\subsection{Trends with AGN Bolometric Luminosity}

\begin{figure}[t]
\includegraphics[width=\columnwidth]{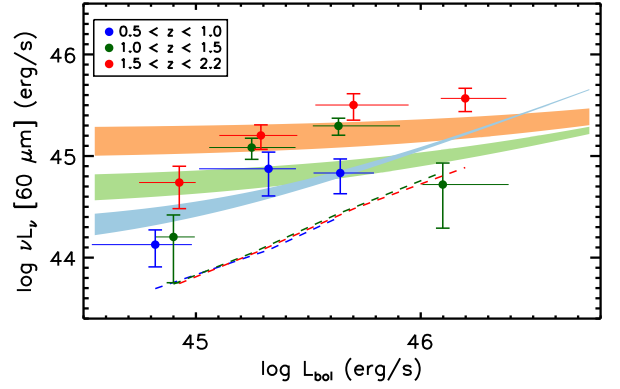}
\caption{The mean \lfir\ of QSOs as a function of AGN bolometric luminosity \lbol, shown as colored circular points with error bars.
Blue/green/red colors represent AGNs in the $0.5<z<1.0$, $1.0<z<1.5$ and $1.5<z<2.2$ redshift 
bins respectively. The X-axis error bars show the range in \lbol\ that encompass 80\%\ of the sample in the corresponding
bolometric luminosity bin.  Dashed lines show the expected contribution to the FIR luminosity from pure AGN-heated dust (Equation 1).
The lightly shaded colored polygons show the mean \lfir\ trends for X-ray AGNs from the fits in \citet{rosario12}.
The difference in the shape of the trends between X-ray AGNs and QSOs
may be attributed to selection biases in the sample (see Section 5.2 for details).
}
\label{lbol_trends}
\end{figure}

In \citet{rosario12}, we explored the relationships between SF and the luminous output of X-ray
AGNs, uncovering a relationship between \lfir\ and \lbol\ in luminous AGNs at $z<1$. This increase suggested an elevated
role for mergers at high AGN luminosities, as predicted by various evolutionary models for SMBH growth \citep[e.g.,][]{hopkins08}.
Here we place our BLAGN sample in the context of our earlier study, to test if the trends uncovered among all X-ray
AGN are also evident in the unobscured luminous AGN studied in this work.

In Figure \ref{lbol_trends}, we plot mean \lfir\ of the BLAGNs, measured from stacks in bins of \lbol\ and redshift.
The AGNs were divided into four bins in log \lbol\ ($44.5-45.0$, $45.0-45.5$, $45.5-46.0$ and $46.0-46.5$) 
as well as the fiducial redshift bins. Additionally, only those AGNs with log \mbh\ between 7.0 and 9.5 were considered
for stacking for consistency with other binning schemes discussed below. The data points in the Figure show 
the mean \lfir\ for all bins with enough objects for a valid measurement ($N>3$). The y-axis error bars come 
from bootstrapping into the stacked sample, while x-axis error bars that show the range in \lbol\ that 
encompass 80\%\ of the subsample in each bin. Included as well in the Figure are lightly
shaded regions which show the empirical trends between \lbol\ and \lfir\ uncovered for X-ray AGNs
from \cite{rosario12}. These trends suggest a correlation between AGN luminosity and SFR at $z<1$, but
a flatter relationship at higher redshift.

The BLAGNs display a different behaviour in the \lbol--\lfir\ plane compared to X-ray AGNs. The FIR luminosities
of the AGNs in the lowest \lbol\ bin are systematically lower than with those of the X-ray AGNs at the same bolometric luminosity. 
In addition, the trends of the BLAGN measurements
in all three redshift bins show a characteristic shape, with a sharp increase in \lfir\  with \lbol\ at low AGN bolometric 
luminosities that flattens or drops at high AGN luminosities in a redshift dependent manner.  At low \lbol, optical BLAGN
appear to show a lower SFR than X-ray AGN of the same bolometric luminosity, while at intermediate \lbol\ ($\sim 10^{45.5}$ \ergs)
they appear to show an elevated mean SFR. A first comparison of the
different trends in the \lbol--\lfir\ plane may lead one to conclude that luminous unobscured AGNs show 
different host SF signatures compared to a broader X-ray selected sample. However,
a closer examination of the selection effects inherent in such BLAGN samples suggests a different interpretation. 

From Figure \ref{mbh_lbol}, we see that more luminous AGN are, on average, associated with more massive SMBHs.
Since SMBH mass is correlated with the stellar mass of the host galaxy, which is in turn correlated with SFR, 
the lower luminosity BLAGNs in our sample contain a larger fraction of low mass host galaxies, which, 
in turn, will bring down their average SFR. At high \lbol, the characteristic flattening seen at low and intermediate
redshift AGNs may be interpreted as either of two ways: a) the most luminous AGN are responsible for quenching
SF in their hosts, or, alternatively, b) such AGN are typically found in high mass galaxies (due to the selection effects
outlined above), which, for various reasons not yet well understood, contain a higher quenched fraction
at all redshifts. As we show in Section 6.1, a model that does not require widespread AGN-driven quenching
in luminous BLAGNs can explain these trends quite well.


Using Equation 1, we estimate the AGN contribution to the 60 \mics\ luminosity of the BLAGNs
in our sample in the different bins in \lbol\ and redshift. For these estimates, we also include a random
term to capture the scatter in $k_{60}$, since our measured mean \lfir, a linear combination, is
disproportionately affected by upward logarithmic scatter than downward scatter. The expected
AGN contribution to \lfir, evaluated from 1000 random estimates, is shown as dashed lines in Figure \ref{lbol_trends}.
Except for the lowest luminosity bin at low redshift and some of the highest luminosity bins, AGN
contamination does not strongly influence trends we see in this diagram, especially those seen at $z\sim2$.
It is also unlikely to influence any of the other trends we present later in this work. 
Accounting for it may, however, strengthen the turnover that we see at high AGN luminosities in the low
and intermediate redshift bins in Figure \ref{lbol_trends}. In Section 6, we develop the discussion of AGN 
contamination and its role in the interpretation of these trends.

\subsection{Trends with Black Hole Mass}

\begin{figure}[t]
\includegraphics[width=\columnwidth]{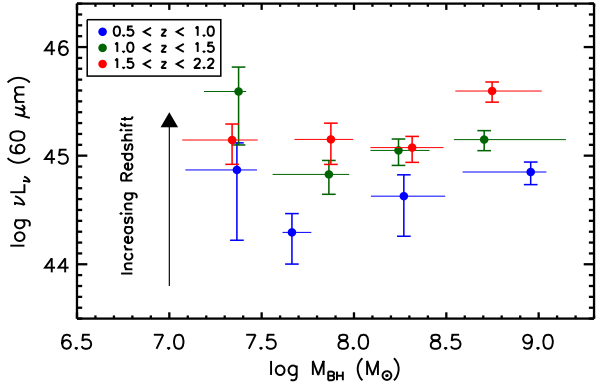}
\caption{The mean \lfir\ of BLAGNs as a function of SMBH mass \mbh, shown as colored circular points with error bars.
Blue/green/red colors represent AGNs in the $0.5<z<1.0$, $1.0<z<1.5$ and $1.5<z<2.2$ redshift bins respectively.
The X-axis error bars show the range in \mbh\ that encompass 80\%\ of the sample in the corresponding
mass bin.
}
\label{mbh_trends}
\end{figure}

Black Hole Mass (\mbh) is a fundamental property of an SMBH and
determines the maximum amount of energy that can be derived from the accreting system. SMBH
demographic studies have revealed a tight correlation between the mass of a black hole and the total
stellar content of the host spheroid in local galaxies \citep[the \mbh--$\sigma$ relation,][]{haring04, sani11},
the form of which may evolve with redshift \citetext{\citealp{jahnke09,merloni10, bennert11}, but see \citealp{schulze11}}. 
Since the spheroid mass is related to the total mass of the galaxy, correlations between \mbh\ and \smass\
are expected and indeed found \citep[e.g.,][]{sani11}. 

The SFR of galaxies is known to correlate with stellar mass along a ridgeline in SFR-\smass\ space
popularly called the `Mass Sequence '  of galaxies  \citep{noeske07, elbaz07, daddi07a, rodighiero11, whitaker12}.
When combined with the \mbh--$\sigma$ relation, the SFR Mass Sequence  predicts a dependence of the 
mean SFR of BLAGNs with \mbh, since more massive SMBHs are found in more massive hosts, 
which in turn have a larger SFR. Do we find such a relationship among our BLAGN sample?

In Figure \ref{mbh_trends}, we plot the mean \lfir\ vs.~\mbh, binning the AGNs in redshift and SMBH mass following the
scheme in Table 1. 
In the low redshift bin, there is a significant increase in the mean SFR between the hosts of  
$10^{7.5} < M_{BH} < 10^{8.0}$ \msun\ black holes and the hosts of $10^{8.5} < M_{BH} < 10^{9.5}$ \msun\ black holes.
The mean SFR of systems with $10^{8.0} < M_{BH} < 10^{8.5}$ \msun\ SMBHs lies in between and spans the difference,
within the errors.
A similar trend, with some scatter, is seen at higher redshifts, with the increase extending to the highest masses probed.
In contrast, there is a hint that the mean FIR luminosities of the lowest mass SMBHs, in the $10^{7.0} < M_{BH} < 10^{7.5}$ \msun\
bins, do not follow the trend exhibited by more massive SMBHs, but, in the low and intermediate redshift bins,
have systematically higher \lfir\ than expected. This may indicate that the lowest mass SMBHs are in hosts with
elevated specific SFRs than the rest of the SMBH population. We caution, however, that the enhancement is
marginal at best and is subject to the highly biased nature of the low mass SMBH population in our sample (Section 4.1).

To constrain the trend between the median \mbh\ (in log \msun) and the mean \lfir\ (in log \ergs), we
fit a simple straight line to the stacked points, excluding the pathological lowest mass bins and including the errors in mean \lfir. 
The slopes of the trends from the fits are $0.71\pm0.47$, $0.34\pm0.20$ and $0.74\pm0.25$ in the low, intermediate
and high redshift bins respectively. Within the uncertainties, the slopes are consistent with being constant with redshift, 
with a weighted average value from all three redshift bins of $\approx 0.5$. This is similar to but a bit shallower than the expected value
of $0.7$ at $z=1$ if SFR $\propto$ \smass$^{0.57}$ \citep{whitaker12} and \smass\ $\propto$ \mbh$^{0.79}$ \citep{sani11}.
The uncertainties in our measurements prevent us from making any conclusions about possible evolution in the relationship
between \mbh\ and SFR, which could arise from, for example, a change in the slope of the \mbh-\smass\ relationship
with cosmic time.

\subsection{Trends with Eddington Ratio}

We now turn to the study of relationships between the SFR of BLAGN hosts and the specific accretion rate
of the SMBH (\redd). In the context of close co-evolution between galaxies and their SMBHs, many models predict strong
evolution in the stellar content of an AGN host galaxy while the SMBH is growing at a fast rate. Therefore, one expects a
higher than average SFR for AGN hosts with fast growing SMBHs, while galaxies with slowly growing SMBHs
will show a slower growth rate or relatively normal SFR for their stellar mass. There is some evidence for this
among local Seyfert 1 AGNs \citep{sani10}.

\begin{figure}[t]
\includegraphics[width=\columnwidth]{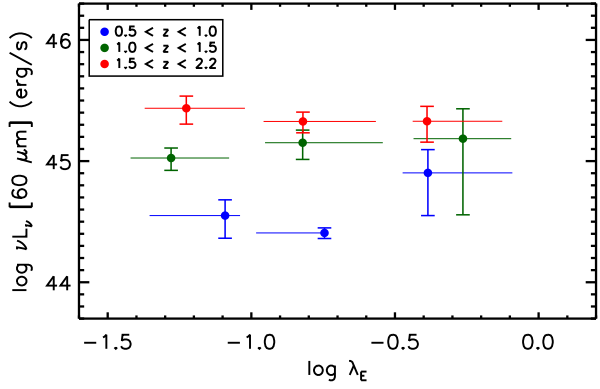}
\caption{ The mean \lfir\ of BLAGNs as a function of AGN specific accretion rate (the Eddington ratio \redd), 
shown as colored circular points with error bars.
Blue/green/red colors represent AGNs in the $0.5<z<1.0$, $1.0<z<1.5$ and $1.5<z<2.2$ redshift bins respectively.
The X-axis error bars show the range in \redd\ that encompass 80\%\ of the sample in the corresponding
Eddington ratio bin. }
\label{ledd_trends}
\end{figure}

In Figure \ref{ledd_trends}, we show measurements of the mean FIR luminosities for our BLAGN 
sample in three bins of redshift and \redd\ (as listed in Table 1). In general, there are no strong systematic trends
between \redd\ and \lfir. Among the two higher redshift subsamples, \lfir\ is consistent with being flat between
\redd\ $=0.03$ and \redd\ $=1$, while in the low redshift bin, there is a increase of \lfir\ in AGNs with \redd\ $> 0.3$
by a factor of 3 over slower accreting objects, significant at the level of $\approx 2\sigma$.

How would the biases inherent in our sample influence these trends? As discussed in Section 4.1 and from Figure \ref{mbh_lbol}, 
the most highly accreting objects at all redshifts include a larger proportion of low mass SMBHs. Low \mbh\ systems
are found in low mass host galaxies, which, in turn, have lower typical SFR. Therefore, the mean SFR of our subsample
of highly accreting SMBHs may be depressed somewhat by the greater fraction of low mass host galaxies compared to
AGNs at lower \redd. If a positive correlation does indeed exist between SFR (or \lfir) and \redd, 
it will be flattened by the inclusion of these low mass galaxies at high \redd. However, a more detailed 
understanding of the effects of selection biases requires modeling, which we pursue in the following section. 


\section{Discussion}

We explored two alternative approaches to mitigate the effects of the sample biases on the observed trends between \lfir\
and various Type-I AGN properties. One was to apply a lower limit in SMBH mass of \mbh$>10^{7.8}$\msun\ to our
sample. As one may see in Figure \ref{mbh_lbol}, this serves to remove the long tail to low \mbh, particularly at $z>1$. 
The resultant reduced sample is less sensitive to \mbh-based selection effects. However, this approach also discards any information 
carried among the lower mass SMBHs, while the smaller sample size leads to insufficient numbers of objects for a good PACS
stacking signal in the important high \redd\ bins. Therefore, we turned to a second approach. We developed a heuristic
empirically-constrained model for the AGN host galaxy population, to enable a prediction of mean \lfir\ and its trends directly from our
measurements of \mbh, based on known scaling relationships between SMBH mass, \smass\ and SFR. By default, 
this subsumes any biases in the sample into the predicted trends. Here, we present the details of this approach, show how
selection effects in the sample influence (and restrict) the results of our mean FIR study, 
and discuss what we may learn about the BLAGN host galaxy population from the trends.

\subsection{The Baseline Model}

In order to make an estimate of the mean \lfir\ for our BLAGNs, we require some input on the nature
of AGN emission, their host galaxies and the evolution of their scaling relations. We make the following assumptions:
a) AGN hosts are subsets of normal star-forming galaxies, i.e., they are not preferentially in `special' populations such as 
starbursts or major mergers; b) AGN hosts lie on the Mass Sequence  of normal star-forming galaxies; 
c) the \mbh-\smass\ relationship remains constant with redshift; 
d) the FIR bolometric correction (Equation 1) of AGNs does not evolve with redshift.
These assumptions define a `baseline model' for the BLAGN population. 
If there are substantial deviations between the data and our predictions, this will serve
as a test of the assumptions built into the baseline model.
 
A prediction of the mean \lfir\ for our BLAGNs is developed as follows. For all objects in a bin of redshift
and/or either \mbh, \lbol\ or \redd, we estimate stellar masses from SMBH masses by inverting Equation 8 of 
\cite{sani11}. From \smass, we derive SFRs using the Mass Sequence relation from \cite{whitaker12}. We convert the 
SFRs to 60 \mics\ luminosities using the standard calibration from \citet{kennicutt98}, taking a mean ratio of 0.5 
between \lfir\ and the integrated 8-1000 \mics\ luminosity, based on the FIR SED libraries of \citet{ce01}. 
We use a Monte-Carlo bootstrap approach to propagate the scatter in these relations -- 
$\sigma$(\mbh/\smass)$=0.35$ dex, $\sigma$(\smass/SFR)$=0.3$ dex and
$\sigma$(SFR/\lfir)$=0.18$ dex -- by randomly varying the assumed relationships around their central trend
by these $\sigma$. In addition, we include a small enhancement to the \lfir\ to account for the AGN hot-dust
contribution at 60 \mics, which is calculated from \lbol\ using Equation 1, with a scatter of 0.3 dex. 
From 1000 iterations, we arrive at a mean predicted SFR for the ensemble of objects in each bin as well as
a prediction for the uncertainty on the mean arising from the intrinsic scatter of AGN host galaxy properties
in our model.

\begin{figure*}[t]
\includegraphics[width=\textwidth]{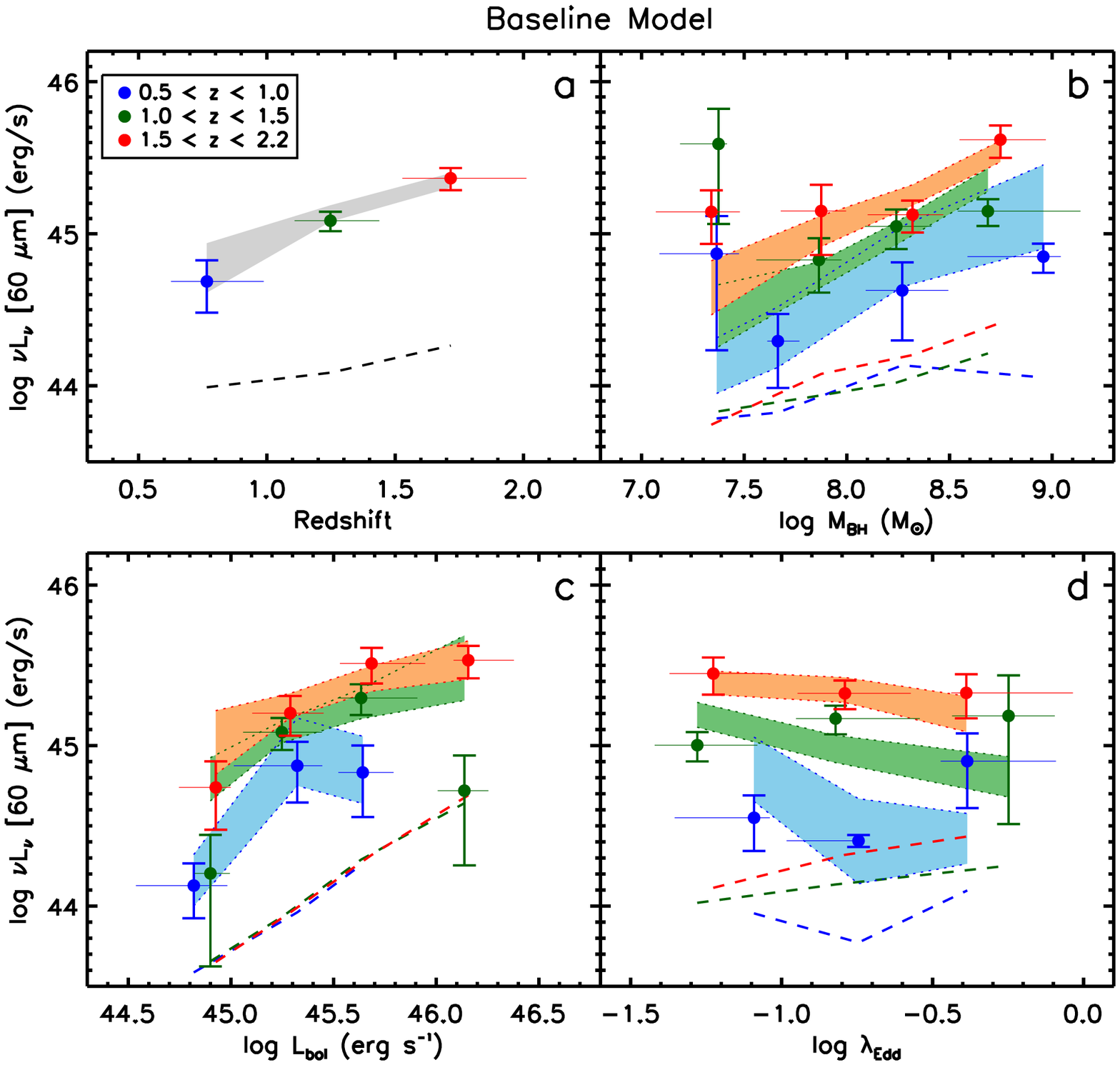}
\caption{A comparison between the measured mean \lfir\ and the predictions of a ``baseline model", which, through a simple
bootstrap analysis, takes into account some of the selection effects built into the QSO sample (see Section 6.1 for details).
Blue/green/red colors represent AGNs in the $0.5<z<1.0$, $1.0<z<1.5$ and $1.5<z<2.2$ redshift bins respectively.
The four panels show measurements as colored points with error bars, in (a) bins of redshift only, (b) bins of \mbh\ and redshift,
(c) bins of \lbol\ and redshift, and (d) bins of \redd\ and redshift. The X-axis error bars are not uncertainties, 
but a range in the abscissa that encompass 80\%\ of the sample in the corresponding bin. 
Dashed lines show the expected contribution to the FIR luminosity from pure AGN-heated dust.
The shaded polygons show the baseline model predictions for the trends: grey for panel (a) and blue/green/red in the other
three panels representing the three redshift bins according to the key in (a). The Y-axis width of the polygons give the $1\sigma$
uncertainty on the mean model trends, determined by the intrinsic scatter of the model QSO host
population and the size of each subsample. In all panels, the form of the trends and the uncertainties on the measurements
are reproduced reasonably well by the baseline model, especially in bins with $>10$ objects.
}
\label{baseline_model}
\end{figure*}

In Figure \ref{baseline_model}, we compare predictions of mean \lfir\ against our \lfir\ measurements. The shaded
regions indicate the range in mean \lfir\ expected from the baseline model following our Monte-Carlo treatment
of scatter. The X-axis values of the model regions are pinned to the actual median value of 
\mbh, \lbol\ and \redd\ in any given redshift bin, since the input to the model are the empirical 
measurements for the very objects that belong to each bin.
In general, the model matches the observed data points quite well, both in the redshift evolution of \lfir,
the trends with SMBH parameters, and also in the scatter about these trends. The biggest
deviations between the data and the model arise in bins with small numbers of objects ($N \lesssim 6$) -- these
bins can be severely affected by stochastic effects which are not accounted for by the bootstrap error estimates. 
The uncertainty on the mean \lfir, while large in these bins, are probably still underestimated.

The PACS stacks indicate a positive correlation between \lfir\ and \mbh. The predicted slope of this trend in the baseline model
is determined almost completely by the slope of the SF Mass Sequence and is a good representation of the 
actual trends seen in the data, indicating that BLAGN hosts also lie along the SF sequence to $z=2.2$. 
As alluded to in Section 5.2, the rather unusual shapes of the \lfir-\lbol\ 
trends are now shown to be driven almost completely by selection effects in the \mbh-\lbol\ plane. 
The baseline model assumes no direct connection between \lfir\ and the accretion luminosity of the AGN 
(except for a small, generally negligible, component from AGN heated dust), yet the predicted
trends show a characteristic slope, which arises only because the low \lbol\ bins contain a larger fraction of low-mass AGN host
galaxies which bring down the mean \lfir. Similarly, while the baseline model does to adopt any explicit relationship between
the SMBH Eddington ratio and SFR, it predicts a flat or falling trend of \lfir\ with \redd\ since the population of
the most rapidly accreting black holes in our sample also include a greater fraction of low-mass systems. 

The good agreement between the baseline model and the stacked \lfir\ measurements suggests that such a simple
model can actually be a good representation of the population of BLAGN host galaxies. Taken at face value,
this agreement implies that BLAGN hosts are not clearly in strong starbursts, since this population is expected to lie well above
the SF Mass Sequence. Instead, most QSO host galaxies are apparently in normal massive star-forming galaxies. However,
we caution against too liberal an interpretation, since this conclusion depends on the validity of the assumptions that
go into the baseline model. Firstly, the model assumes that all host galaxies lie on the SF Mass Sequence, i.e., AGNs
hosts are all forming stars. It is known, however, that a substantial fraction of massive galaxies lie in quiescent hosts, which we
have not accounted for in the model. Optical imaging of QSO hosts suggest that most show signs of active on-going SF 
\cite[e.g.,][]{jahnke04, trump13}, while Herschel-based studies of lower luminosity X-ray AGNs show that they
are preferentially found in SF galaxies \citep{rosario13}. Taken together, the assumption that QSOs are in SF hosts
may be reasonably valid. Besides this, objects with very weak 
or absent SF do not contribute significantly to the stacked flux, meaning that possible quiescent AGN hosts in our sample
are deweighted in the mean measured \lfir. Another wrinkle arises because 
the adopted \mbh\-\smass\ relationship \citep[Equation 8 of][]{sani11}
is valid only for the bulge stellar mass, while, in the fashion of \citet{merloni10}, we use it with the total galaxy stellar mass.
In addition, we are also aware that SMBH scaling laws may vary at the highest mass 
end (\mbh$> 10^9$ \msun) and lowest mass end (\mbh$< 10^7$ \msun) 
\citep[e.g.,][]{graham12,vdbosch12,graham13}, but this is unlikely to influence our results, 
since most of our AGNs have masses within these extremes.

The sizable uncertainties on the measurements prevent a finer investigation into
the parameter space of the models. Given the primarily empirical nature of this work, we restrict our study of models
to a few simple tests, in which we explore the performance of the models when we vary the offset of the AGNs from the
SF Mass Sequence, or include possible evolution in the \mbh-\smass\ scaling relation.

\begin{figure*}[t]
\includegraphics[width=\textwidth]{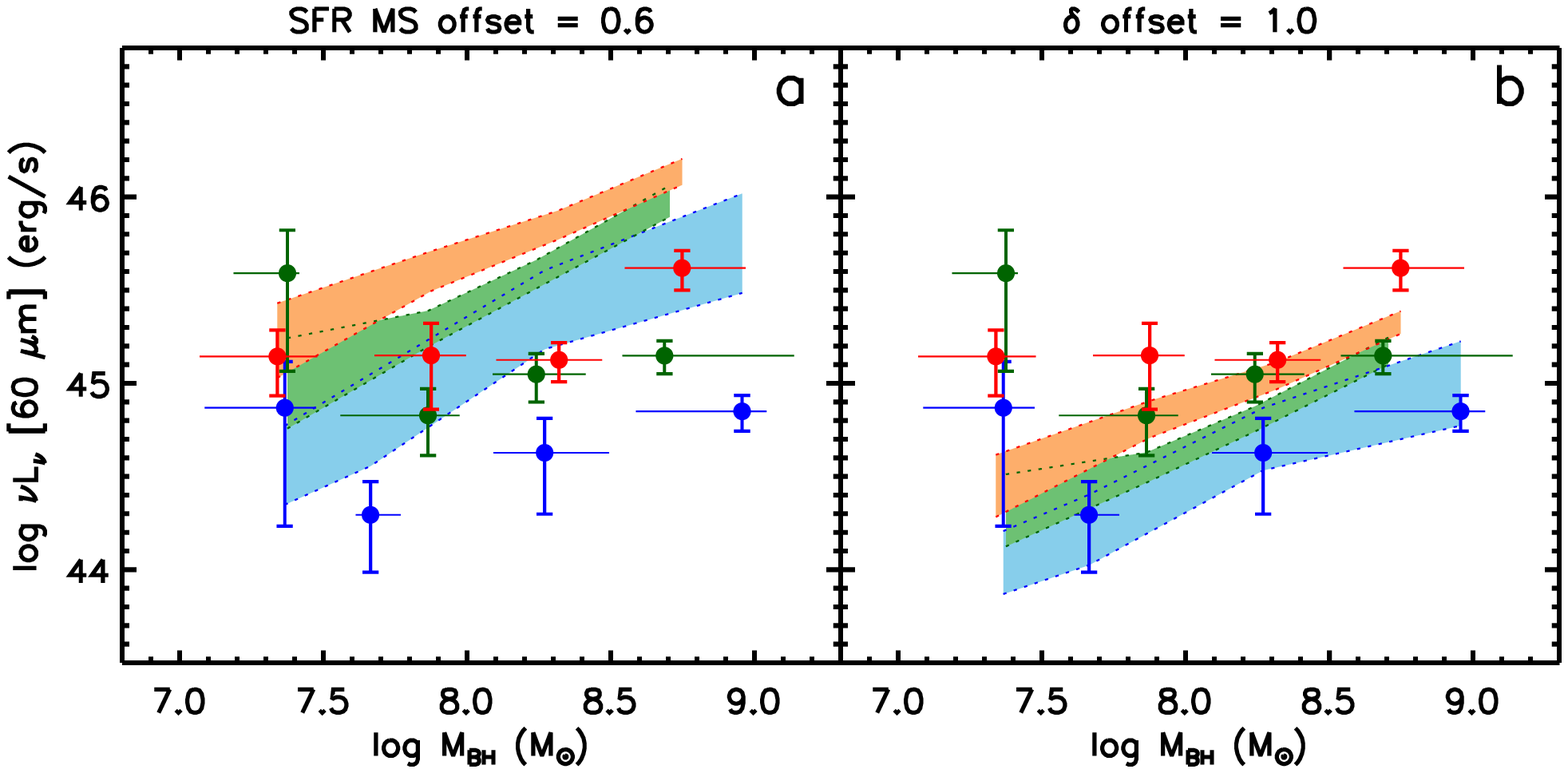}
\caption{A comparison between the measured mean \lfir\ in bins of \mbh\ and the predictions of two simple alternative
models for BLAGN host properties. (a) All BLAGNs are assumed to lie in starbursts
which have at a mean positive offset of 0.6 dex in SFR from the SF Mass Sequence, or (b) the \mbh-\smass\
relation evolves with redshift according to Equation 2 with $\delta=1.0$.
The measurements are shown as colored points with error bars.
Blue/green/red colors represent AGNs in the $0.5<z<1.0$, $1.0<z<1.5$ and $1.5<z<2.2$ redshift bins respectively.
The X-axis error bars show the range in the \mbh\ that encompass 80\%\ of the sample in the corresponding bin. 
Neither set of parameters chosen here fit the data as well as the `baseline model' (Section 6.1).
While model (a) performs very poorly in fitting the data, model (b) is marginally acceptable, though smaller values
of $\delta$ are preferred.
}
\label{modelvars}
\end{figure*}

\subsection{Varying the offset from the SF Mass Sequence}

Co-evolutionary scenarios that link AGN activity with bursts in SF predict that AGNs should be found in host galaxies
that lie above the SF Mass Sequence. We perform a simple test of these scenarios by adding a redshift-independent 
offset to the SFR in our baseline model and then compare the goodness-of-fit of the altered model to that of the baseline
model (i.e., the model with normal star-forming AGN hosts). As a measure of the goodness-of-fit, we calculate
the $\chi^{2}$ statistic of the model predictions with respect to the stacked measurements in bins of
\mbh\ and redshift bins (Figure \ref{mbh_trends}). The bootstrap errors on the measurements
are used to weigh the data in the estimation of the statistic. Better fitting models yield lower values of $\chi^{2}$.
We caution, however, that the $\chi^{2}$ discussed below are meant to allow a qualitative judgement of 
variation arising from model parameters. One should not use them for robust statistical estimates of
confidence intervals or for significance testing.

For the baseline model, we calculate a $\chi^{2}=3.1$. If AGN hosts are in a starburst phase offset in SFR by +0.6 dex 
from the Mass Sequence \citep{rodighiero11, sargent12}, the $\chi^{2}$ increases to 7.5 (left panel of Figure \ref{modelvars}). 
This sharp increase clearly disfavors a scenario where all BLAGNs are in strong starbursts. Even a minor SFR offset 
over the Mass Sequence results in an increased $\chi^{2}$ (to 3.4 for a +0.1 dex offset, for example). Indeed, the
$\chi^{2}$ reaches a minimum with \emph{negative} offsets, placing QSOs very slightly \emph{below} the 
Mass Sequence (-0.05 dex gives a $\chi^{2}=3.0$). 
Since there is some debate as to the exact form for the Mass Sequence \citep{noeske07, elbaz07, daddi07a, wuyts11, whitaker12},
it may be that the baseline model is a slightly incorrect representation of the true galaxy population, and certainly a -0.05 dex
offset is within most uncertainties in the Sequence. Another explanation arises
if we consider evolution in the \mbh-\smass\ relation from the canonical form we adopted in the baseline model.

\subsection{Varying the normalization of the \mbh-\smass\ relation}

Recent studies of the evolution of SMBH scaling laws suggest an increase in the normalization of the
\mbh-\smass\ with redshift \citep[e.g.,][]{decarli10, merloni10, trakhtenbrot10, bennert11, targett12}, which can be
parameterized in the following fashion:

\begin{equation}
\Delta M_{BH}/M_{*} \; = \; \delta \log (1+z)
\end{equation}

\noindent where $\Delta M_{BH}/M_{*}$ is the offset from the local \mbh-\smass\ relation \citep{sani11}. These various studies
have calibrated the slope of the redshift term $\delta =0.5{\rm -}2.0$, with substantial uncertainty. 
A positive $\delta$ results in a smaller \smass\ for a given \mbh\ at higher redshifts, 
which will produce a smaller mean SFR and lower \lfir\ compared to the baseline model. 
However, reviews of the biases inherent in studies of the evolution of SMBH scaling laws \citep{schulze11, salviander13}
suggest that $\delta$
may be rather unconstrained by empirical studies and even $\delta=0$ could be consistent with such studies. 
By including a term from Eqn.~2 to
the \mbh-\smass\ relation in the baseline model, we can test for the effects of changing SMBH relations. We
find that a small positive value of $\delta$ does improve the performance of the model, but only slightly
($\delta=0.25$ yields a minimum $\chi^{2}=3.0$). Stronger redshift evolution leads again to larger $\chi^{2}$ -- for e.g., 
at $\delta = 0.5$, the low end of the range from direct studies, $\chi^{2}=3.1$, while for $\delta = 2.0$, $\chi^{2}=4.8$.
In the right panel of Figure \ref{modelvars}, we show the performance of the model with $\delta = 1.0$ ($\chi^{2}=3.4$).
Clearly, our measurements are most consistent with very mild to no redshift evolution in the \mbh-\smass\ relation.

\subsection{The Star Forming Properties of QSO host galaxies}

We have shown that the mean SFR of QSOs with nuclear bolometric luminosities in the range \lbol$=10^{44.5\textrm{--}46.5}$
can be described quite well by our `baseline model', in which QSO hosts are normal SF galaxies which lie on the Mass Sequence.
We find that a slight evolution in the \mbh-\smass\ relation with redshift is supported by the modeling of our measurements.
We also show that the scenario in which QSO hosts contain starbursts (i.e., have a positive offset from the SF Mass Sequence)
and the scenario where \mbh\ increasingly precedes \smass\ with redshift (Eqn.~2 with a positive $\delta$) have opposite
effects on the modeled SFR-\mbh\ relationship. Therefore, a situation in which QSO hosts, in truth, lie increasingly among starbursts 
at higher redshifts may be offset by a positive evolution in the \mbh-\smass\ relationship. Satisfying a pure starburst scenario
\citep[as defined by][]{rodighiero11, sargent12} is highly unlikely, since it would require a very high $\delta$ which is not supported
by observations. At present, our measurements cannot distinguish 
between the simple baseline model and a scenario with modest $\delta$ evolution coupled with a moderate starburst fraction.
Having said this, the simplest scenario, based on the assumptions of the baseline model, works quite well and a more complex model
will have to bring together firm complementary evidence to support it over the basic baseline model.

\section{Conclusions}

Combining diverse spectroscopic datasets from the COSMOS extragalactic survey, we compile one of the 
largest samples of moderate luminosity QSOs to $z\sim2$ with deep FIR data from the PEP survey
and uniform measurements of SMBH properties. We extensively characterize the sample and highlight important
selection effects which play a role in understanding its properties. The QSO database is used to explore the relationships
between SFR and \mbh, \lbol\ and SMBH specific accretion rate \redd. After accounting for selection effects
using a Monte-Carlo bootstrapping procedure, we show that the SFR trends of QSOs out to $z\sim2$ are most consistent with a simple
model where their hosts are galaxies that lie on the SF Mass Sequence. Scenarios where all QSO hosts
are in strong starbursts are inconsistent with our measurements. The typical SFRs of galaxies hosting the fastest
growing black holes are not significantly enhanced over systems with slower growing black holes.
Our modeling also indicates that the redshift 
evolution of SMBH-host scaling relationships may be rather mild. Taken at face value, our results suggest that QSOs at the luminosities
that dominate the volume-averaged SMBH growth at $z=2$ lie in fairly normal star-forming host galaxies, which set important
constraints on models of AGN-galaxy co-evolution and the processes that influence SMBH scaling laws.

\acknowledgements

PACS has been developed by a consortium of institutes led by MPE (Germany) and including UVIE
(Austria); KU Leuven, CSL, IMEC (Belgium); CEA, LAM (France); MPIA (Germany); INAF-IFSI/
OAA/OAP/OAT, LENS, SISSA (Italy); IAC (Spain). This development has been supported by the
funding agencies BMVIT (Austria), ESA-PRODEX (Belgium), CEA/CNES (France), DLR (Germany),
ASI/INAF (Italy), and CICYT/MCYT (Spain). This research has made use of the NASA/IPAC 
Infrared Science Archive, which is operated by the Jet Propulsion Laboratory, 
California Institute of Technology, under contract with the NASA. zCOSMOS is based on observations 
made with ESO Telescopes at the La Silla or Paranal Observatories under programme ID 175.A-0839.

\bibliographystyle{aa}

\bibliography{cosmos_type1}

\end{document}